\documentclass[superscriptaddress, nofootinbib, preprintnumbers, twocolumn]{revtex4}

\usepackage{amsmath}
\usepackage{amsthm}		
\usepackage{amssymb}	
\usepackage{bm}
\usepackage{xcolor}
\usepackage{datetime}
\usepackage{eufrak}
\usepackage{graphicx}
\usepackage{setspace}
\usepackage{verbatim}
\usepackage[hidelinks]{hyperref}

\definecolor{grey}{rgb}{0.4,0.4,0.4}
\definecolor{dullmagenta}{rgb}{0.4,0,0.4}
\definecolor{darkblue}{rgb}{0,0,0.4}
\definecolor{midblue}{rgb}{0,0,0.5}
\definecolor{midred}{rgb}{0.5,0,0}
\definecolor{orange}{rgb}{1,0.5,0}
\definecolor{lightbrown}{rgb}{0.75,0.5,0.25}
\definecolor{tan}{cmyk}{0.14,0.42,0.56,0}
\definecolor{djunglegreen}{cmyk}{0.99,0,0.52,0}
\definecolor{lightgreen}{rgb}{0,1,0}
\definecolor{olivegreen}{cmyk}{0.64,0,0.95,0.40}
\definecolor{midgreen}{rgb}{0.0,0.675,0.0}
\definecolor{darkgreen}{rgb}{0,0.5,0}
\definecolor{LightOrange}{rgb}{0.98,0.68,0.24}
\definecolor{LightChameleon}{rgb}{0.54,0.88,0.20}
\definecolor{LightScarletRed}{rgb}{0.93,0.16,0.16}
\definecolor{DarkSkyBlue}{rgb}{0.12,0.29,0.53}

\newcommand{\vs}{\vspace}

\newcommand{\Prm}{\ensuremath{\mathrm{P}}}

\newcommand{\grm}{\ensuremath{\mathrm{g}}}

\newcommand{\srm}{\ensuremath{\mathrm{s}}}

\newcommand{\PBH}{{\rm PBH}}

\smallskipamount

\settimeformat{ampmtime}

\usepackage{float}

\begin{document}

 \title{New Mass Window for Primordial Black Holes as Dark Matter \\
 from  Memory Burden Effect}

\author{Ana Alexandre}
\email{alexand@mpp.mpg.de}
\affiliation{
	Arnold Sommerfeld Center,
	Ludwig-Maximilians-Universit{\"a}t,
	Theresienstra{\ss}e 37,
	80333 M{\"u}nchen,
	Germany,}
\affiliation{
	Max-Planck-Institut f{\"u}r Physik,
	F{\"o}hringer Ring 6,
	80805 M{\"u}nchen,
	Germany}

\author{Gia Dvali}
\affiliation{
	Arnold Sommerfeld Center,
	Ludwig-Maximilians-Universit{\"a}t,
	Theresienstra{\ss}e 37,
	80333 M{\"u}nchen,
	Germany,}
\affiliation{
	Max-Planck-Institut f{\"u}r Physik,
	F{\"o}hringer Ring 6,
	80805 M{\"u}nchen,
	Germany}

\author{Emmanouil Koutsangelas}
\email{emi@mpp.mpg.de}
\affiliation{
	Arnold Sommerfeld Center,
	Ludwig-Maximilians-Universit{\"a}t,
	Theresienstra{\ss}e 37,
	80333 M{\"u}nchen,
	Germany,}
\affiliation{
	Max-Planck-Institut f{\"u}r Physik,
	F{\"o}hringer Ring 6,
	80805 M{\"u}nchen,
	Germany}

\date{\formatdate{\day}{\month}{\year}, \currenttime}

\begin{abstract}

\vspace{0.2cm}

\noindent 

The mass ranges allowed for Primordial Black Holes (PBHs) to constitute all of Dark Matter (DM) are broadly constrained. However, these constraints rely on the standard semiclassical approximation which assumes that the evaporation process is self-similar.  Quantum effects such as memory burden take the evaporation process out of the semiclassical  regime latest by half-decay time.  What happens beyond this time is currently not known.  However, theoretical evidence based on prototype models indicates that the evaporation slows down thereby extending the lifetime of a black hole.  This modifies the mass ranges constrained, in particular, by BBN and CMB spectral distortions. We show that previous constraints are largely relaxed when the PBH lifetime is extended, making it possible for PBHs to constitute all of DM in previously excluded mass ranges. In particular, this is the case for 
PBHs lighter than $10^9$ g which enter the memory burden stage before 
BBN and are still present today as DM.  

\end{abstract}
\maketitle

\section{Introduction}
\label{sec:Introduction}
\vs{-5mm}

Since it was first proposed that black holes could form from primordial fluctuations \cite{Zeldovich_PBH_1967,Hawking_PBH_1971,Hawking_Carr_PBH_1974}, the possibility of Primordial Black Holes (PBH) constituting all or part of the Dark Matter (DM) in the Universe has been considered \cite{Chapline:1975ojl}. This possibility has recently seen a renewed interest as a result of the LIGO detection of merging black hole binaries with masses around $1-50 M_{\odot}$ \cite{LIGOScientific:2018jsj} whose formation is not easily explained by astrophysical processes.
Although at the moment there is no observational evidence for their existence, several constraints have been put on the fraction of DM in the form of PBHs \cite{CarrFlorianDM, CKSY,CKSY2}, defined as 

\begin{equation}
    f_{PBH} (t) \equiv \frac{\rho_{PBH} (t)}{\rho_{DM} (t)},
\end{equation}

\noindent for different values of the PBH mass $M$ in terms of physical time $t$. Currently, there are only a few mass ranges of interest that still leave the door open for $f_{PBH} =1$ 
(for a review on PBHs, see \cite{Carr_kuhnel_PBH_review}).\\

Another important point of interest is that PBHs are the only ones that can be small enough for Hawking radiation to be relevant \cite{Hawking:1974rv}.
For a PBH forming from primordial fluctuations, its mass should be comparable to the horizon mass at the time of its formation, $t_f$. Assuming this takes place during radiation domination, we can estimate it as
    \begin{equation}
        M
            \sim 
                \rho
                V
                \bigg|_{t_f}
            \sim 
                \frac{M_\Prm^2}{H}
                \bigg|_{t_f}
            \sim 
                M_\Prm t_f
            \; .
    \end{equation}
Here, $\rho$ denotes the total energy density of the universe, $H$, the Hubble parameter and, $V(t) \sim H(t)^{-3}$, the Hubble volume.\\

The standard Hawking evaporation time, which describes the lifetime of a PBH of mass $M$, can be expressed as
    \begin{equation}
    \label{Eq:HawkingEvap}
        t_H
            \sim 
            r_g \, S
            \sim
                \frac{M^3}{M_\Prm^4}
            \; ,
    \end{equation}
where 
    \begin{equation}
        r_g 
            \sim 
                \frac{M}{M_\Prm^2}
            \; ,
        \qquad
        S
            \sim 
                \left(
                    \frac{M}{M_\Prm}
                \right)^2,
            \; 
    \end{equation}

\noindent denote the gravitational radius and entropy of the black hole, respectively. In this scenario, for a PBH to be present in the Universe today, its mass must be larger than $M \sim 10^{14}$g, meaning that smaller PBHs cannot account for any DM.\\

However, this estimate is based on the semi-classical evaporation scenario which relies on the assumption of self-similarity. 
That is, during its evaporation, a black hole gradually shrinks in size while maintaining the standard semi-classical relations between its parameters, such as its mass, the radius and the temperature.  
Notice that this is a very strong assumption, since  the evaporation rate has been reliably derived exclusively for a black hole of a fixed radius and in exact-zero back-reaction limit:  there exists no reliable semi-classical calculation that can take into account the effects of quantum back-reaction on the black hole geometry over a long time-scale. 
Despite this,  self-similarity is a standard assumption in all the current  
derivations of constraints on PBHs. 
More recent findings show that the above assumption is not only unjustified but it is actually inconsistent over the long time-scales of the evolution. That is, the semi-classical approximation cannot hold throughout the entirety of the black hole lifetime.\\
  
The first evidence of this came from a microscopic picture which resolves a classical black hole as a
coherent state of gravitons \cite{Dvali:2011aa, Dvali:2012en, Dvali:2013eja}.
This picture shows that, due to the quantum back-reaction, any such state must go out of the validity of the semi-classical approximation 
latest by the time of its half-decay, $t_{half}$.
This time marks the point when a black hole emits approximately half of its initial mass, and should not be confused with its half-lifetime. In particular, as we shall explain, the second stage of the decay can last much longer.\\

The strength of the quantum back-reaction is of order  $\sim 1/S$ per emission time\footnote{As shown in \cite{Dvali:2015aja}, the deviation from Hawking of order $1/S$ per each emission must be present independently of a microscopic theory. This is very important for two reasons. Firstly, it is much larger than the standard instanton-type effects that would be expected to scale as $e^{-S}$. Secondly, although for large black holes the $1/S$-correction still appears to be small, one must remember that it is inversely proportional to the number of emissions during $t_{half}$ which scales as $\sim S$. As a result, the cumulative effect over half-decay time is of order one, irrespective of the initial mass and entropy of the black hole.} and comes from two effects that go hand in hand. The first one is the loss of coherence generated by the inner entanglement among the constituent  gravitons \cite{Dvali:2013eja}.
Note that this process is fundamentally different from the process of entanglement between the black hole state and its decay products, originally discussed by Page \cite{Page:1993wv}. It therefore must be taken into account separately.\\

The second engine, the most relevant for the present paper, is the so-called ``memory burden" effect \cite{Dvali:2018xpy, Dvali_BHmetamorphosis}. 
A powerful property of this effect is that it goes beyond a particular microscopic description of a black hole and is generic for systems of high microstate entropy  $S$. Such systems possess an exponentially large number, $e^S$, of degenerate microstates which can store large amounts of quantum information. Due to this, systems with $S \gg 1$ exhibit an enhanced capacity of information storage.\\

In any such system, including a black hole, the microstate degeneracy is due to a set of nearly-gapless modes, the so-called ``memory modes".  The number of distinct species of memory modes is $\sim S$.  The degenerate microstates are formed by various excitation patterns of these modes, which encode the information carried by the system.
The high information capacity is due to the large number and low energy cost of the memory modes.
During the decay of the system, the mode-degeneracy gets lifted and the information patterns become more and more costly in energy. This generates a back-reaction that works against the decay and is the essence of the memory burden effect
\cite{Dvali:2018xpy, Dvali_BHmetamorphosis}.
\\ 
   
In the case of a black hole, the inevitability of the memory burden effect can be understood from the following simple reasoning \cite{Dvali:2018xpy, Dvali_BHmetamorphosis}. 
If we assume that the evaporation of a black hole is self similar, then, by definition, after the time $t_{half}$, the black hole must shrink to half of its original radius.
Correspondingly, the entropy must become a quarter of its original entropy. That is, under the assumption of a self-similar evaporation, by the time $t_{half}$, the parameters have to evolve in the following way:
      \begin{equation} \label{TH}
     M  \rightarrow \frac{M}{2}\,,~\, 
        r_g 
            \rightarrow \frac{r_g}{2} \,,~ \, 
         S 
            \rightarrow \frac{S}{4} \,. 
           \end{equation}

Now, we must take into account that, by the very same assumption of self-similarity, during this time, the information cannot be released efficiently: 
self-similarity implies that at each stage radiation is thermal and information is maintained internally. Thus, information must remain encoded in the black hole. However, the remaining black hole has only a quarter of its initial entropy and therefore a much less information storage capacity.
This leads to the following: 
Firstly, under no circumstances can such a black hole be described semi-classically as it would carry much more information than its entropy.  
Secondly, the information pattern can no longer be encoded in gapless modes and thus becomes very costly in energy, creating the memory burden effect. \\

The memory burden effect was studied in detail both numerically and analytically on prototype systems of high-information storage capacity \cite{Dvali_BHmetamorphosis}. In all cases, it was confirmed that the back-reaction reaches its maximum the latest by $t_{half}$. Due to a generic nature of the phenomenon, as well as the model-independent arguments of the type displayed above, the stabilizing tendency 
of memory burden is shared by black holes with high likelihood.\\
 
The key question for PBH DM is:  what happens after $t_{half}$?  
The definite answer to this question is not known. However, we can rely on the following results to formulate our guidelines. 
The analysis of prototype models clearly shows that after $t_{half}$ the system gets effectively stabilized and gains a much longer lifetime \cite{Dvali_BHmetamorphosis}.
However, this result must be taken with a grain of salt since beyond $t_{half}$ the universality of the evolution is no longer guaranteed, and different systems may progress differently.
Correspondingly, extending the results obtained in the analysis of prototype systems to black holes requires some guess-work.\\

In general, as discussed in \cite{Dvali_BHmetamorphosis}, two options stand out.
The first possible scenario is that, once the memory burden reaches its maximum, a black hole develops a new collective (classical) instability, not captured by the prototype analysis. Due to this, a black hole may experience a sharp transition to some lower energy state or even disintegrate. We shall not pursue this path here.\\

An alternative scenario is that the memory burden effect continues to stabilize the black hole. In this case, the lifetime of a black hole is expected to be prolonged substantially.
As pointed out in \cite{Dvali_BHmetamorphosis}, one implication of this effect would be the opening of a new mass region for PBH DM. This is natural, since, in the new circumstances, PBHs much lighter than $10^{14}$g can survive until today. Some steps of exploration of the new potential mass range were already taken in \cite{Dvali_BHmetamorphosis}.
In the present paper we shall continue this line of research.\\

Nonetheless, there exist other constraints on these lighter PBHs, beyond their evaporation. In particular, one must consider those arising from Big Bang Nucleosynthesis (BBN) and the Cosmic Microwave Background (CMB) \cite{Carr:2009jm}. In the case of BBN, the final abundance of heavy elements is highly dependent on the nuclei density before and during nucleosynthesis. If the PBH density is too large, the number and energy density of emitted particles can substantially change the outcome of the processes required, therefore, changing the final abundance of heavy elements \cite{CKSY2}.\\

Indeed, this has been a topic of interest since the early period of PBH research. The precise effect on the consecutive interactions that lead to the formation of primordial nuclei depends on the emitted particles. For example, Vainer and Nasel'skii studied how the injection of high-energy neutrinos and antineutrinos from PBHs would change the weak-interaction freeze-out, thus changing the neutron-to-proton ratio at the start of BBN leading to an overproduction of Helium-4 \cite{Vainer_Naselski_1977,Vainer_Naselski_1979}. Moreover, the increase of the entropy-to-baryon ratio during nucleosynthesis by evaporating PBHs can also lead to the overproduction of Helium-4 and underproduction of Deuterium \cite{Miyama_Sato_78}.
Another important analysis is when Zel'dovich \textit{et al} studied the effects of neutron-antineutron production during nucleosynthesis, arguing that the capture of free neutrons by protons and the spallation of Helium-4 would increase the deuterium abundance \cite{Starobinskii_PBH_BBN}. This was later confirmed numerically by Vainer \textit{et al} \cite{Vainer_1978}.
Furthermore, Lindley discussed the dissociation of deuterons by emitted photons \cite{Lindley_1980} and found it comparable to the limits derived by Zel'dovich in \cite{CKSY}.\\

It is important to mention that the further development of numerical methods in the consequent decades allowed for much more complete analysis of the effects of evaporating PBHs on BBN. In particular, the precision of observational data on the neutron lifetime and primordial abundances have increased and the understanding of the hadronization of the emitted quark and gluon jets has significantly grown. Long-lived hadrons, i.e., those whose lifetime is at least comparable  to the timescale of nucleosynthesis, can leave a significant imprint on BBN, when emitted by both low \cite{Kohri_1999} and high-mass \cite{CKSY} PBHs.
Since all the above mentioned bounds have been calculated using the ordinary Hawking evaporation rate, they must all be appropriately recalculated.\\

In this work, we revise the constraints from BBN and CMB on $f_{PBH}$, in the region with $M \lesssim 10^{14}$g, assuming a suppressed evaporation rate motivated by the studies 
of the memory burden effect \cite{Dvali_BHmetamorphosis}. Our goal is to identify the unconstrained range where PBHs could account for all of DM.
By estimating the emitted energy density, we can conclude that the suppressed emission of PBHs during these periods results in a reduced contribution to these physical processes, which relaxes the bounds dramatically. We rewrite both the BBN and CMB bounds in terms of a generalized lifetime, leading again to the conclusion that both these bounds are relaxed by memory burden.\\

Before proceeding, we would also like to acknowledge the work in progress 
by \textbf{Valentin Thoss} et al \cite{Valentin2024}, which also studies the implications of memory burden effect on PBH DM.\\
\section{Memory burden and black hole evaporation rate}
\label{sec:Memory_Burden}

 As already discussed, we shall adopt the picture in which, after its half-decay, the black hole enters the memory burden phase and continues with a suppressed emission rate. Following \cite{Dvali_BHmetamorphosis}, we will assume that evaporation is slowed down by further $n$ powers of the entropy $S$
 so that the remaining lifetime becomes, 
    \begin{equation}
    \label{Eq:ModEvap}
        t^{(n)}
            \sim 
                S^{1+n}\,r_g
            \; ,
    \end{equation}
where $n$ is a non-negative integer. Although non-integer $n$ could also be considered, we justify the present choice by assuming that the decay rate is analytic in $S$.\\

The case with $n=0$ corresponds to the standard picture in which the semi-classical regime is extrapolated until the end of the black hole lifetime. The regime with $n=1$ corresponds to the minimal suppression after half-decay. The modified lifetime of a black hole is thus
    \begin{equation}
    \label{Eq:BHLifetime}
        t_H^{(n)}
            \sim    
                \frac{M^{3+2n}}{M_\Prm^{4+2n}}
            \; .
    \end{equation}

Although in reality the memory burden effect increases gradually, within the validity of the order-of-magnitude estimate, we shall adopt a simplified picture. 
Namely, we will consider the evolution of a PBH as if it were divided into two stages. In the first stage, its evaporation rate is 
described by the ordinary Hawking formula. This stage lasts until the PBH has lost about half of its mass. 
Beyond this time, the lifetime is described by Eq.~(\ref{Eq:ModEvap}) with $n>0$.\\

As  was adopted in \cite{Dvali_BHmetamorphosis} it is reasonable to expect that the radiation after the memory burden stabilization must continue to be soft, i.e., the emitted quanta on average must not exceed energies of order $1/r_g$. This is natural, since by the time of memory burden, the growing energy gaps of the memory modes (which initially are $\sim 1/(Sr_g)$)
do not exceed $\sim 1/r_g$ \cite{Dvali:2021bsy}. Thus, the black hole continues to be composed of mostly soft quanta and the radiation must accordingly be soft.\\

\section{Modified parameter space for PBHs as DM}
\label{sec:PBH_DM_range_n_estimate}

According to Eq.~(\ref{Eq:ModEvap}), the mass of PBHs evaporating at $t_H^{(n)}$ is given by
    \begin{equation}\label{M_n}
        M^{(n)}
            \sim
                \left(
                    M_\Prm^{4 + 2n} t_H^{(n)}
                \right)^\frac{1}{3 + 2n}
            \sim
                10^{-6} \grm  
                \left(
                    \frac{t_H^{(n)}}{10^{-42} \srm}
                \right)^\frac{1}{3+2n}
            \; .
    \end{equation}

\noindent This equation tells us that, in the case of standard Hawking evaporation, only PBHs of $M^{(0)} \gtrsim 10^{14}$ g are present for the current age of the universe, $t_0 \sim 10^{17}$s. In contrast, PBHs of mass $M^{(1)} \gtrsim 10^6$ g would be present today when we consider the minimal case of modified evaporation. Thus, a new viable window is opened for PBH as DM in the range $10^6 \ \grm \lesssim M^{(1)} \lesssim 10^{14}\grm$.\\


In order to understand how the BBN and CMB constraints are modified by the memory-burden-induced evaporation stage, we first identify which mass ranges are affected. 
For any given epoch with temperature $T$, novelty only concerns the PBHs that enter the second (memory-burden-dominated) evaporation stage at some higher temperature $T_M > T$, that is, the PBHs that satisfy

\begin{equation} \label{tBBNtH}
    t_{half} <  H^{-1} \bigg|_{T} \,.
\end{equation}

\noindent Equivalently, using the fact that $t_{half} \sim M^3/M_P^4$ and $H^{-1} \sim M_P/T^2$ (assuming radiation domination), we can rewrite (\ref{tBBNtH}) in terms of the PBH mass, 

\begin{equation} \label{MofBBN}
    M < M_P \left (\frac{M_P}{T} \right )^{\frac{2}{3}} \,. 
\end{equation}

In principle, any period of the Universe's history sensitive to evaporating black holes can be used to derive a bound on the standard parameter $\alpha$, the fraction of the Universe mass in PBHs, defined as
    \begin{equation}
        \alpha
            \equiv 
                \frac{\rho_\PBH}{\rho_m}
                \bigg\vert_{t_0}
            \; ,
    \label{Eq:PBHFractionToday}
    \end{equation}

\noindent where $\rho_m$ is the total energy density of matter and radiation. Such constraints come from the requirement that the evaporation does not inject too much energy into the existing radiation bath of the Universe. In order to calculate them, let us first estimate the energy density emitted during this period.\\

Taking into account the additional suppression in \eqref{M_n}, the black hole mass can be estimated to evolve in time as
    \begin{equation}
    \label{mass_time_evolution}
        \dot{M}    
            \sim 
                - M_{\Prm}^2
                \left( 
                    \frac{M_{\Prm}}{M} 
                \right)^{2+2n}
            \; .
    \end{equation}
    
\noindent During a given epoch with temperature $T$ ending at $\Delta t$, such a black hole is emitting the energy

\begin{align}
       \Delta E
            &=
                | \dot{M} \Delta t|\sim \Delta t M_\Prm^2 \left(\frac{M_\Prm}{M}\right)^{2+2n}\; .
    \label{Eq:EmittenEperBH}
\end{align}

\noindent For example, we can take $\Delta t_{\text{BBN}} M_P^2 \sim 10^{72}$eV for BBN. It can be seen that the black hole is injecting very energetic photons but with a density much smaller than the photon density at that time.\\

Assuming PBHs are formed during radiation-domination with a monochromatic mass spectrum, their number density at temperature $T$ can be expressed as 
    \begin{align} 
        n_{\rm PBH} (T) 
            &\sim 
                \frac{\rho_{\rm PBH}(T)}{M}
            \sim 
                \frac{\rho_{\rm PBH}(t_0)}{M}
                \left(
                    \frac{T}{T_0}
                \right)^3
            \, .
    \label{Eq:PBHNumberDens}
    \end{align}
 Therefore, the total amount of emitted energy is approximately
    \begin{align}
        \Delta E_{\rm tot}(H^{-1})
            &=
                (\# BH) \Delta E
                \bigg|_{T} 
            \; ,   
    \label{Eq:TotEmittedEnergy}
    \end{align}

\noindent and the radiation injected by the PBHs into their surroundings per unit time can be approximated in terms of the current energy density of PBHs as at most 

\begin{equation}
   \dot{\rho}_{PBH} (T) \, \sim 
 \frac{\rho_{PBH} (t_0) }{t_H^{(n)}} \frac{T^3}{T_0^3} \,.  
\end{equation}

Any physical process happening in this period will be unaffected, provided the injection is negligible when compared to the radiation energy density due to the Hubble expansion,  

\begin{equation}
   |\dot{\rho}_{H}| \, \sim \, T^4 \, H\, \sim 
   \frac{T^{6}}{M_P}\,.  
\end{equation}

\noindent Assuming that PBHs constitute all of DM, this condition can be quantified by the following parameter

\begin{equation} \label{EPS}
  \epsilon(T) \equiv  \frac{ |\dot{\rho}_{PBH}|}{|\dot{\rho}_{H}|} 
   \sim   \frac{\rho_{DM}}{t_{H}^{(n)}} \frac{1}{T_0^3 T H}  \,.
\end{equation}

Taking the fact that $T_M$ is approximately the temperature at $t_{half} \sim M^3/M_P^4 \sim M_P/T_M^2$, we can rewrite the expression for $t_{H}^{(n)}$ in terms of $T_M$, so that $\epsilon(T)$ becomes, 

\begin{equation} \label{EPSn}
  \epsilon (T) \, 
  \sim \,     \left ( \frac{\rho_{DM}}{T_0^3 T} \right) 
   \left ( \frac{M_P}{H} \right) 
   \left (\frac{T_M}{M_P} \right )^{2+\frac{4}{3}n}
 \,.
\end{equation}

\noindent We can apply the above to various epochs by evaluating $\epsilon$ at the corresponding temperature $T$ such that PBHs in the mass-range of interest are unconstrained if $\epsilon (T) \ll 1$.
Since $\epsilon(T) \propto T^{-3}$, with all other parameters fixed, it becomes more significant for lower $T$. The most stringent constraint therefore comes from the present epoch,  $T=T_0$, which using (\ref{EPSn}) can be translated as a condition on $n$. 
For example, for $T_M \sim 10$MeV,  we have $\epsilon(T_0) \sim 10^{21 -28n}$, which is already negligible for $n=1$.
Note that the power of this argument is that it holds regardless of the nature of the injected particles or which specific physical processes are affected by the evaporation.\\

\section{Rewriting BBN bounds with an extended PBH lifetime}
\label{sec:BBN_bounds}

In the case of BBN, $T = T_{\rm BBN} \sim 1$MeV, the PBHs obeying \eqref{tBBNtH} have mass in the range  

\begin{equation} \label{MofBBN}
    M^{(1)} < M_P \left (\frac{M_P}{T_{BBN}} \right )^{\frac{2}{3}} \,
    \sim 10^{14}M_P \sim 10^9 {\rm g}\,. 
\end{equation}

\noindent Therefore, the region with $M \gtrsim 10^9$ g remains unaffected and the standard BBN constraints still apply.
However, PBHs in the range $10^6 \ \grm \lesssim M^{(1)} \lesssim 10^{9} \grm$ are not only present today but have also entered their second evolution stage before or during BBN. This is the range of interest for the present study where constraints are modified.\\

For such black holes, taking into account that $t_{H}^{(n)}  >  t_0$, we can estimate an upper bound on $\epsilon (T_{\rm BBN})$ as 

\begin{equation}
  \epsilon(T_{BBN})     
    <    \frac{M_P}{t_0}  \frac{\rho_{DM}}{T_0^3 T^{3}_{BBN}} \,
 \sim  10^{-25}  \,. 
\end{equation}

\noindent Clearly, PBHs in the mass-range of interest are completely unconstrained by BBN.\\ 

However, to connect with the existing literature on BBN contraints mentioned in section \ref{sec:Introduction}, one can further estimate for some particular cases.
For example, if we assume that the emitted energy is entirely in the form of photons, then 

\begin{equation}
\Delta E_{\rm tot}(t_{\rm BBN})
            \simeq
                \left(
                    \frac{n_{\rm PBH}}{n_\gamma} \right) \Delta E \bigg|_{t_{\rm BBN}} 
            \; .
\end{equation}

\noindent Moreover, the total emitted energy must be below the energy of the background photons, i.e.,  $\Delta E_{\rm tot}(t_{\rm BBN}) \lesssim E_\gamma(t_{\rm BBN}) \sim T_{\rm BBN}$. Using (\ref{Eq:PBHNumberDens})-(\ref{Eq:TotEmittedEnergy}) yields

    \begin{equation}
        \rho_{\rm PBH}(M) 
            \lesssim 
                \frac{\rho_\gamma (t_{\rm BBN})}{M_\Prm \Delta t_{\rm BBN}}
                \left(
                    \frac{M}{M_\Prm}
                \right)^{3+2n}
                \left(
                    \frac{T_0}{T_{\rm BBN}}
                \right)^3
            \; .
    \end{equation}

We can express this as a constraint on $\alpha$ with (\ref{Eq:PBHFractionToday}). Moreover, for a crude estimate, using $\rho_\gamma(t_{\rm BBN}) \sim T_{\rm BBN}^4$ and $\rho_m(t_0) \sim 10^{3} T_0^4$, we find

    \begin{equation}
        \alpha
            \lesssim 
                \frac{10^{-3}}{M_\Prm \Delta t_{\rm BBN}}
                \left(
                    \frac{M}{M_\Prm}
                \right)^{3+2n}
                \left(
                    \frac{T_{\rm BBN}}{T_0}
                \right)
            \; .
    \end{equation}

\noindent The emitted PBH energy during BBN is too small a fraction of the photon energy for PBHs of this initial mass to have a large enough impact that could further constrain the PBH mass spectrum.\\

Alternatively, we can also consider the case where all the energy gets converted into neutrons as in \cite{Starobinskii_PBH_BBN}. We can estimate the fraction of produced neutrons over the total neutron number density as $\Delta n_B / n_B \sim 10^{-37.5}$. Clearly, we cannot expect PBHs to produce enough neutrons to affect the final product of nucleosynthesis. We thus would expect the bounds on PBHs as DM from BBN to be relaxed if the evaporation is slowed down.\\

Nevertheless, it is worth for completeness to attempt to rewrite existing bounds in terms of the parameter $n$ in order to verify their consistency with the argument presented in the previous section. For this purpose, it turns out convenient to express the constraint in terms of another standard parameter $\beta$, which stands for the fraction of the Universe’s mass in PBHs at the time of their formation,
    \begin{equation}
        \beta
            \equiv 
                \frac{\rho_\PBH}{\rho_m}
                \bigg\vert_{t_{\rm f}}
            \; .
    \label{Eq:PBHFractionFormation}
    \end{equation}
Notice that, at $t_{\rm f}$, the PBHs form in the radiation-era, meaning that radiation is taken into account in $\rho_m$.
This ratio is related to $\alpha$ by
    \begin{equation}
        \beta
            =
                \alpha
                \left(
                    \frac{M_\Prm}{M}
                \right)^{1+n}
            \; .
    \end{equation}

The ratio at the time of the PBH formation, $\beta$, should be independent of their evaporation rate, since the black hole formation occurs before its evaporation.
This does not mean of course that $\beta$ cannot be affected by considering the changes in the evaporation process on constraints from today's observations (see section \ref{sec:CMB_distortion_bounds}).
The same, however, is not true of $\alpha$ or the ratio $\beta / \alpha$, which is given in terms of the initial Primordial Black Hole mass by
    \begin{equation}
    \label{beta-alpha}
        \frac{\beta}{\alpha} 
            = 
                \left( 
                    \frac{M_P}{M} 
                \right)^{1+n}
            \; .
    \end{equation}
A relation between $\alpha(n=0) = \alpha_0$ and $\alpha(n>0) = \alpha_n$ in terms of the evaporation time can then be deduced as 
    \begin{equation}
    \label{alpha_alpha0}
        \alpha_n (t^{(n)}_H) 
            = 
                \alpha_0 (t^{(0)}_H) \frac{(t^{(0)}_H M_P)^{-1/3}}{(t^{(n)}_H M_P)^{-\frac{1+n}{3+2n}}} 
            = 
                \alpha_0 (M) 
                \left( 
                    \frac{M}{M_P} 
                \right)^n
            \; .
    \end{equation}
Note that $\alpha_n (t^{(n)}_H)$ should not be confused with evaluating $\alpha$ at $t_H^{(n)}$; it is still evaluated at $t_0$, and is simply taking into account the extended evaporation time.
Equation (\ref{alpha_alpha0}) can be used to rewrite any existing estimates of BBN bounds on the PBH mass function in a form that more generally accommodates for a slower evaporation time.\\

As an example, this can be applied to the case of the injection of neutrons by PBHs, following the estimates in \cite{Starobinskii_PBH_BBN}. In particular, for black holes evaporating during BBN, the bound becomes
    \begin{equation}
    \label{alpha_B1}
        \alpha_n 
            \leq 
                10^{10.5 + 39n} 
                \left( 
                    \frac{M}{M_{\odot}} 
                \right)^{\frac{1}{2} + n} 
                \Omega
            \; .
    \end{equation}

\begin{figure}
    \centering
    \includegraphics[width=\linewidth]{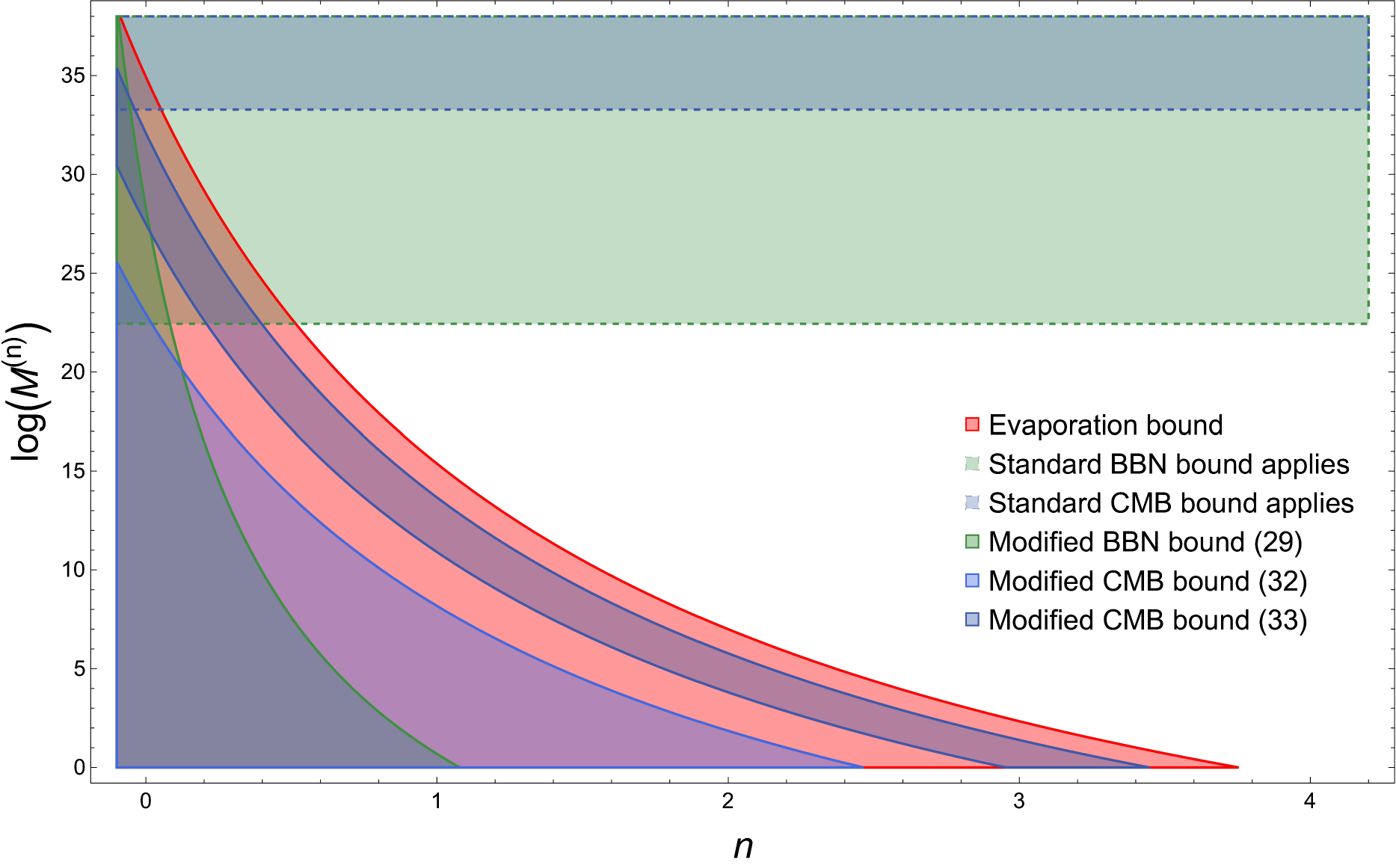}
    \caption{Plot of the logarithm of initial PBH masses $M^{(n)}$ allowed in terms of the memory-burden-strength parameter $n$. The dashed regions represent the mass ranges where the usual constraints apply for BBN (green) and CMB spectral distortions (blue). It can be seen that the modified bounds for both BBN (green) and (CMB) only apply for masses already excluded by the evaporation bound (red). The unshaded region (white) represents the new window that is open for PBH DM, which grows together with the parameter $n$.}
    \label{Fig:M_n}
\end{figure}

Requiring that $\alpha_n \leq \Omega$ implies that, for $n=1$, $M^{(1)} \leq 1$g, which is much smaller than the minimum initial mass for the PBHs to be present at BBN, $M^{(1)} \geq 10^4$g. That is, the mass is constrained only in a range that is outside of the relevant mass range. Therefore, the bound coming from nucleosynthesis is actually fully relaxed for $n=1$, as we would expect from the estimate carried out in section \ref{sec:PBH_DM_range_n_estimate}. The same applies for the case of $n=2$, where the bound becomes $M^{(2)} \leq 10^{-4.4}$g but only black holes of initial mass $M^{(2)} \geq 1$g would be present. This can be seen in Figure \ref{Fig:M_n}, where the initial PBH mass $M^{(n)}$ is plotted against the parameter $n$ with a logarithmic scale. The green shaded region with the solid boundary representing the bound in equation \eqref{alpha_B1} is completely inside the red shaded region, representing the evaporation bound.
That is, for any $n$, the bound only applies to masses already constrained by evaporation.
It can be seen further that the higher the parameter $n$ is, i.e., the stronger the memory burden affect is, the larger the new window for PBH DM becomes.\\

We would like to briefly comment that the accretion effects are unlikely to be significant for light PBH DM in the new mass window. 
For example, for PBHs entering the memory burden phase before BBN, the geometric cross section is $<10^{-38}$cm$^{2}$.
If placed in the center of a nucleon (proton or neutron), the gravitational potential of PBH on the scale of the nucleon size is $\sim 10^{-5}$ of the QCD potential. Similarly, the induced gravitational shift of the atomic levels is of order $10^{-6}$.\\

\section{Rewriting CMB-distortion bounds with an extended PBH lifetime}
\label{sec:CMB_distortion_bounds}

As mentioned in section \ref{sec:Introduction}, the mass range that becomes accessible to PBHs as DM is also constrained by CMB spectral distortions, as previously discussed in \cite{Carr:2009jm}.
Following the same line of reasoning as in the nucleosynthesis section \ref{sec:BBN_bounds}, for $T = T_{\rm CMB} \sim 3000$K\footnote{Note that $T_{\rm CMB}$ refers to the temperature at which the CMB was emitted, not its current measured value.}, the highest mass of interest is given by 

\begin{equation} \label{MofBBN}
    M^{(1)} < M_P \left (\frac{M_P}{T_{CMB}} \right )^{\frac{2}{3}} \,
    \sim 10^{18}M_P \sim 10^{13} {\rm g}\,. 
\end{equation}

\noindent Therefore, PBHs in the range $10^6 \ \grm \lesssim M^{(1)} \lesssim 10^{13} \grm$ are present today and have also entered their second evolution stage before or during the emission of the CMB. We can estimate how constraints are altered as the lifetime of the black holes is extended in this mass range.\\

For such black holes, taking into account that $t_{H}^{(n)}  >  t_0$, we can estimate an upper bound on $\epsilon (T_{\rm CMB})$ as 

\begin{equation}
  \epsilon(T_{CMB})     
    <    \frac{M_P}{t_0}  \frac{\rho_{DM}}{T_0^3 T^{3}_{CMB}} \,
 \sim  10^{-5}  \,. 
\end{equation}

\noindent Although this is considerably larger than in the nucleosynthesis case, it is still small enough that we expect PBHs in the memory burden phase evaporating during CMB emission to be mostly unconstrained.\\

Once again, for completeness, let us see how existing constraints would change in terms of $n$.
The first constraint applies to lower mass PBHs that would have evaporated completely before the emission of the CMB. If these emit photons early enough, they are fully thermalized before recombination and contribute only to the photon-to-baryon ratio $\eta$. Therefore, the requirement is that $\alpha (M) < \eta^{-1} \sim 10^9$ and can be rewritten in terms of the parameter $n$ as

\begin{equation} \label{CMB_bound_1}
    \beta (M) < 10^9 \left( \frac{M_P}{M} \right)^{1 + n}.
\end{equation}

\noindent In the case of $n=0$, this applies only to PBHs evaporating around $100$s after the Big Bang with mass below $10^9$g, which corresponds to $M^{(1)} < 10^{3.5}$g.\\

The second bound that appears in this region is for PBHs evaporating during and after the CMB, whose emitted photons, although partly thermalized, can produce significant distortions in the CMB spectrum unless $\alpha (M) < 1$. In terms of the generalized $n$, this translates to the bound

\begin{equation} \label{CMB_bound_2}
    \beta (M) < \left( \frac{M_P}{M} \right)^{1 + n},
\end{equation} 

\noindent which is much stronger than the bound in (\ref{CMB_bound_1}). In the case of standard Hawking evaporation, this bound has been derived in \cite{Carr:2009jm} for the mass range $10^{11}$g$ < M < 10^{13}$g. However, unlike in the previous bound, the translation of this mass range to higher $n$ is no longer trivial since the derivation includes assumptions about the evaporation process. Regardless, we can estimate that the bound applies to PBHs evaporating just before the CMB up to the lowest masses that would still be present today. In the case of $n=1$, this would correspond to $10^{4.5} \textrm{g} < M^{(1)} < 10^{6}$g and, for $n = 2$, it would be $10^{1.5} \textrm{g} < M^{(2)} < 10^{2.5}$g. For the mass ranges in between, the constraint should transition from the weaker bound to the stronger one.\\

We can see from both expressions (\ref{CMB_bound_1}) and (\ref{CMB_bound_2}) that the bound on $\beta (M)$ is lower as $n$ increases. The first impression may be that this implies PBHs are strongly constrained when $n > 0$. However, we should note again that the weaker bound (\ref{CMB_bound_1}) only applies to PBHs that could not constitute DM as they have already evaporated long before the CMB. Moreover, the stronger bound only potentially applies to the lowest masses of PBHs still present today. In fact, one can additionally note that the mass range to which the bound (\ref{CMB_bound_2}) applies becomes narrower as $n$ increases such that more mass ranges are unconstrained overall.\\

This can be seen more clearly in Figure \ref{Fig:M_n}. The blue shaded regions with the solid boundaries representing the two bounds above apply only in the regions already excluded by evaporation, similarly to the nucleosynthesis case. Once again, it can be seen that the stronger the memory burden effect is, the larger the new window for PBH DM becomes.
Analogously to the comment in section \ref{sec:BBN_bounds}, accretion effects are also not expected to be significant in the case of CMB constraints.\\

\section{Role of Species} 

The possible existence of a large number of particle species can significantly affect the physics of black holes
\cite{Dvali:2007hz, Dvali:2007wp, Dvali:2008fd, Dvali:2008ec}. 
 First,  in a theory with $N$ particle species, the size of the smallest  Eisnteinian black hole is bounded from below by the following length scale, the species length
\cite{Dvali:2007hz, Dvali:2007wp},
 \begin{equation} 
  L_{sp} \equiv \frac{\sqrt{N}}{M_P}, 
\end{equation}  
  and correspondingly, the mass of a black hole is bounded by 
  \begin{equation} 
  M_{\rm min}   =  \sqrt{N} M_P \, .
\end{equation}  

  For $t  < t_{half}$, the decay of an Einsteinian black hole 
  must be democratic in species \cite{Dvali:2007hz, Dvali:2007wp, Dvali:2008fd}. 
  This equality can also be understood  
  from the no-hair property \cite{Ruffini:1971bza, Bekenstein:1972ny, Teitelboim:1972pk, Hartle:1971qq} of a semi-classical black hole.  This feature demands that all the species are treated 
  equally by a black hole. In the opposite case, a black hole would acquire a new label, 
  a so-called ``species hair",  according to its bias with respect to different species \cite{Dvali:2008fd}.\\
  
The rate of decay is thereby increased $N$-fold. Correspondingly, the half-decay time from (\ref{Eq:HawkingEvap}) is shortened to

\begin{equation} \label{TN}
  t_{half} = r_g \frac{S}{N} \, .
\end{equation}

\noindent This shortening  applies equally to the meaning of $t_{half}$ as the memory burden time \cite{Dvali:2021bsy}.   
That is, after $t_{half}$ a PBH looses half of its mass and also reaches the maximal memory burden effect. \\

For $t> t_{half}$ the situation is  more subtle. Since the black hole is no longer semi-classical,  the species democracy is not guaranteed from this point of view. 
In other words,  a black hole after $t_{half}$ can develop a ``species hair" which can create a bias in the evaporation rate in favor of some particular species \cite{Dvali:2008fd}. 
However, it is reasonable to assume that during $t> t_{half}$ the evaporation rate  remains democratic in species at least order-of-magnitude wise. 
In this case, the black hole lifetime for $t > t_{half}$ (\ref{Eq:BHLifetime})  will accordingly get shortened by an extra factor $\sim 1/N$.\\
  
This $N$-dependence must be taken into account when applying the analysis of the present paper to theories with large $N$. Notice that in the Standard Model $N \sim 100$.  Among these species, only the ones lighter than $1/r_g$ must be counted. However, in theories beyond the Standard Model, the number of species can be much larger with the upper bound set by $N\sim 10^{32}$ \cite{Dvali:2007hz, Dvali:2007wp}.   
Such extensions however must be considered on a case by case basis, as the increased $N$ can come with other modifications of Einstein gravity, such as large extra dimensions \cite{Arkani-Hamed:1998jmv}.
\\
    
\section{Conclusion \& outlook}
\label{sec:Conclusion_and_Outlook}

The standard lower bound of $10^{14}$g 
on PBH masses (for a review see, \cite{Carr_kuhnel_PBH_review}) is based on the assumption of the self-similarity of the black hole evaporation process.  However, more recent studies based on both microscopic descriptions  \cite{Dvali:2011aa, Dvali:2013eja} and on certain universal aspects of the behavior of systems of maximal microstate entropy \cite{Dvali:2018xpy, Dvali_BHmetamorphosis} paint a very different picture. These studies show that the self-similar regime must fully break down latest by the time $t_{half}$ when the black hole emits around half of its mass.\\ 
 
One of the main contributors to this breakdown is the ``memory burden" effect \cite{Dvali:2018xpy, Dvali_BHmetamorphosis}: the back-reaction from the initially-gapless modes (memory modes) that account for the microstate degeneracy of the system.
This back-reaction resists the decay and becomes strong by $t_{half}$.\\

What happens beyond this point is currently unknown.
However, the analysis of prototype systems performed in \cite{Dvali_BHmetamorphosis} indicates that the decay slows down dramatically.  
It was therefore suggested that, due to the universality of the phenomenon, the same behavior must be exhibited by a black hole, hence extending its lifetime. One of the implications of the effect suggested in \cite{Dvali_BHmetamorphosis} is the lowering of the lower mass bound of PBH DM.
When the lifetime of PBHs is extended, lower mass black holes are no longer constrained by evaporation. 
In the present paper, we extended the analysis of \cite{Dvali_BHmetamorphosis} to study the remaining potential constraints on light PBH DM.\\

An interesting consequence is that the PBHs that survived until today and were present during BBN and CMB emission, not only have lower masses but also decay with a suppressed emission rate. This of course reduces the potential impact they would have had on these periods leading to more relaxed constraints.
This has interesting implications because it allows PBHs to constitute all of DM in lower mass ranges, for example, $10^6$g$ < M^{(1)} < 10^{9}$g, even when we only extend the lifetime by a single power of the entropy. As one can see in Figure \ref{Fig:M_n}, if we extend the lifetime even further, this new window for PBH DM becomes even wider.
Furthermore, this shines light on the fact that a better understanding of black hole physics beyond semiclassicality is crucial to the study of PBHs as a potential DM candidate.\\

\appendix

\acknowledgments

We would like to thank Florian Kühnel for the many helpful discussions on PBHs and their constraints. We also thank Valentin Thoss for lively discussions and communication. 
This work was supported in part by the Humboldt Foundation under the Humboldt Professorship Award, by the European Research Council Gravities Horizon Grant AO number: 850 173-6, by the Deutsche Forschungsgemeinschaft (DFG, German Research Foundation) under Germany’s Excellence Strategy - EXC-2111 - 390814868, Germany’s Excellence Strategy under Excellence Cluster Origins 
EXC 2094 – 390783311. \\

Disclaimer: Funded by the European Union. Views and opinions expressed are however those of the authors only and do not necessarily reflect those of the European Union or European Research Council. Neither the European Union nor the granting authority can be held responsible for them.\\

\setlength{\bibsep}{5pt}
\setstretch{1}
\bibliographystyle{utphys}
\bibliography{Paper_Draft/main}

\end{document}